\documentclass[aip,pop,reprint]{revtex4-1}

\usepackage{amsmath,amssymb,amsthm}

\usepackage{color,graphicx}

\begin{document}


\title{Unique Topological Characterization of Braided Magnetic Fields} 



\author{A. R. Yeates}
 \email{anthony.yeates@durham.ac.uk}
 \affiliation{Department of Mathematical Sciences, Durham University, Durham, DH1 3LE, UK}

\author{G. Hornig}
 \email{gunnar@maths.dur.ac.uk}
\affiliation{Division of Mathematics, University of Dundee, Dundee, DD1 4HN, UK}


\date{\today}

\begin{abstract}
We introduce a topological flux function to quantify the topology of magnetic braids: non-zero, line-tied magnetic fields whose field lines all connect between two boundaries. This scalar function is an ideal invariant defined on a cross-section of the magnetic field, and measures the average poloidal magnetic flux around any given field line, or the average pairwise crossing number between a given field line and all others. Moreover, its integral over the cross-section yields the relative magnetic helicity. Using the fact that the flux function is also an action in the Hamiltonian formulation of the field line equations, we prove that it uniquely characterizes the field line mapping and hence the magnetic topology.
\end{abstract}


\maketitle 


\section{Introduction}

In many plasmas, ranging from astrophysics to magnetic confinement fusion, the topology---i.e., the linking and connectivity of the magnetic field lines \cite{longcope2005}---is an approximate invariant of the dynamics. This is because these plasmas typically have such low dissipation that, to first approximation, their evolution is ideal. That is, on large scales where the magnetohydrodynamic approximation holds, they satisfy an ideal Ohm's law \cite{goedbloed2004}, preserving the magnetic topology. Therefore a practical question is, given two magnetic fields satisfying the same conditions on the boundary of some volume $V$, can one be reached from the other by some ideal evolution in $V$?

We restrict our attention to line-tied magnetic flux tubes, where all field lines stretch between two boundaries and the magnetic field in the volume is non-vanishing. This models, for example, a coronal loop in the Sun's atmosphere, where the footpoints remain essentially fixed on the rapid timescale of coronal relaxation \citep[e.g.,][]{parker1972,*vanballegooijen1985,*mikic1989,*longcope1994,*galsgaard1996, *browning2008,pontin2010}. To simplify the presentation in this paper, we consider a magnetic field ${\bf B}$ defined on a cylinder \mbox{$V=\{(r,\phi,z) | 0\leq r\leq R, 0\leq z\leq 1 \}$}, satisfying $B_z>0$ everywhere in $V$, and impose the boundary conditions that ${\bf B}|_{\partial V}={\bf e}_z$ and ${\bf v}|_{\partial V}=0$, where ${\bf v}$ is the velocity. Extensions of the results to more general boundary conditions are discussed in Section \ref{sec:conclusion}. For convenience, we call magnetic fields satisfying the above conditions ``magnetic braids'' (Fig. 1). Two magnetic braids are topologically equivalent if they can be linked by an ideal evolution where ${\bf v}=0$ on $\partial V$ throughout.

In principle, one can determine whether two magnetic braids are topologically equivalent by comparing their field line mappings from $z=0$ to $z=1$. Throughout this paper, let $f(x_0; z)\in V$ denote the point at height $z$ on the field line traced from $x_0\equiv(r_0,\phi_0,0)$ on the $z=0$ boundary. Under this parameterization of field lines by $z$, we have
\begin{equation}
\frac{\mathrm{d} f(x_0; z)}{\mathrm{d}z} = \frac{{\bf B}\big(f(x_0; z) \big)}{B_z\big(f(x_0; z) \big)}.
\end{equation}
For shorthand, we shall denote the mapping from $z=0$ to $z=1$ as $F(x_0)\equiv f(x_0;1)$.
Under our boundary conditions, two magnetic braids ${\bf B}$, $\widetilde{\bf B}$ are equivalent if and only if $F=\widetilde{F}$. Note that if we were to relax the condition that ${\bf B}={\bf e}_z$ on the side boundary $r=R$, then $F$ would determine the topology only up to an overall rigid rotation through $2n\pi$, $n\in\mathbb{Z}$. Mathematically, field line mappings are symplectic, since they preserve magnetic flux. Symplectic mappings have long been used themselves as models of periodic magnetic fields in fusion devices \cite{morrison2000,*caldas2012}, and field line mappings have also been used extensively for characterizing line-tied coronal magnetic fields \cite{titov2002,*titov2007,*yeates2012,*pariat2012}. But the mapping is usually very sensitive to small fluctuations in the underlying magnetic field. This makes it very difficult, if not impossible, to determine whether two field line mappings can be related by an ideal evolution in anything other than highly idealized situations.

A much more robust topological quantity is the total magnetic helicity, which has a broad range of applications in both laboratory and astrophysical plasmas \cite{brown1999}. It has proved so robust that it has been hypothesized to be the only quantity determining the final state of turbulent relaxation in reversed-field pinches and similar devices \cite{taylor1974,*taylor2000}. But the helicity is an extreme reduction of the topological information in the three-dimensional magnetic field to a single number, and it does not uniquely characterize the topology. There are a large class of magnetic fields that have the same helicity but different field line mappings.

In this paper, we describe a quantitative measure ${\cal A}(x_0)$ which we call the ``topological flux function''. This is a scalar function defined on a cross section of the magnetic field, or equivalently on each field line. It is more robust than the field line mapping, while containing more detailed information than the helicity. A similar function was introduced for magnetic fields in a half-space by Berger \cite{berger1988}, who showed that it is effectively a helicity per field line. The function ${\cal A}$ has also appeared in the literature in a different guise: as an action integral yielding the magnetic field line equations in a variational formalism \cite{cary1983,morrison2000,petrisor2002,*apte2006}. These interpretations are discussed in more detail in Section \ref{sec:interp}.
Recently, we have described how ${\cal A}$ may also be viewed as a generalization of the scalar flux function used to define two-dimensional magnetic fields, and how it may be used to define and measure magnetic reconnection \cite{yeates2011}.
Here we go further. Our main result is that, with a particular choice in  its definition, ${\cal A}$ uniquely characterizes the topology of a magnetic braid. In other words, ${\cal A}=\widetilde{\cal A}$ for two magnetic braids if \emph{and only if} they are topologically equivalent. Moreover, not only can ${\cal A}$ determine whether two braids are topologically equivalent, it can also quantify how much dissipation or reconnection is needed to connect the two states. This will be invaluable in future dynamical studies of such systems.

This paper is organized as follows. In Section \ref{sec:def}, we define ${\cal A}$ and give its basic physical interpretation in terms of magnetic flux. Its interpretations as a field line helicity, as an average crossing number, and as a Hamiltonian action are described in Section \ref{sec:interp}. In Section \ref{sec:main} the Hamiltonian interpretation is used to prove our main result that ${\cal A}$ uniquely characterizes the magnetic topology. We outline in Section \ref{sec:conclusion} how the boundary conditions may be relaxed.

\section{Definition of the Topological Flux Function} \label{sec:def}

The topological flux function ${\cal A}$ is defined simply by integrating the vector potential ${\bf A}$ (where ${\bf B}=\nabla\times{\bf A}$) along magnetic field lines. It may be written as
\begin{equation}
{\cal A}(x_0) \equiv \int_{z=0}^{z=1}{\bf A}\cdot\,\mathrm{d}{\bf l} \equiv \int_0^1\frac{{\bf A}\big(f(x_0;z)\big)\cdot{\bf B}\big(f(x_0;z)\big)}{B_z\big(f(x_0;z)\big)}\,\mathrm{d}z.
\label{eqn:adef}
\end{equation}

\begin{figure}
\includegraphics[width=\columnwidth]{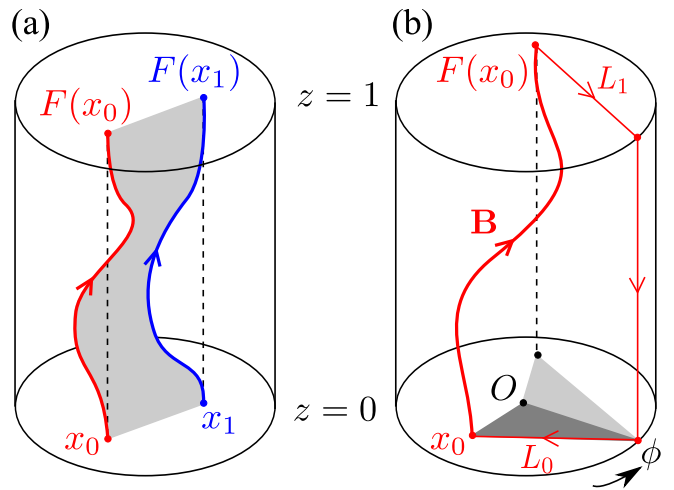}%
\caption{(Color online) Magnetic braids in the cylinder $V$, showing (a) the comoving surface defined by two periodic field lines, and (b) the poloidal surface at viewing angle $\phi$ for a single field line (bounded by the arrowed path).\label{fig:sketch}}
\end{figure}

Broadly, ${\cal A}$ is conceived to measure poloidal (horizontal) magnetic fluxes in the domain $V$. But it will be beneficial to restrict the gauge of the vector potential ${\bf A}$ in the definition \eqref{eqn:adef}. This is because the value ${\cal A}(x_0)$ for a general field line is not gauge invariant. Indeed, under a gauge transformation ${\bf A}\rightarrow{\bf A} + \nabla\chi$,
\begin{equation}
{\cal A}\rightarrow {\cal A} + F^*\chi - \chi,
\label{eqn:gauge}
\end{equation}
where we use the pull-back notation $F^*\chi(x_0)\equiv \chi\big(F(x_0) \big)$. Provided that $\chi$ is chosen to be periodic in $z$, then it follows that for periodic field lines, where $F(x_0)=x_0$, the value of ${\cal A}(x_0)$ is gauge invariant. In an ideal evolution, two periodic field lines define a comoving surface with flux  ${\cal A}(x_1)-{\cal A}(x_0)$, as in Fig. \ref{fig:sketch}(a), and the set of periodic field lines can be used to define a poloidal flux partition of the magnetic field \cite{yeates2011}. But for a general, non-periodic field line, ${\cal A}$ is not gauge invariant unless we impose further gauge conditions on ${\bf A}$ in its definition \eqref{eqn:adef}.

We show here that ${\cal A}$ is a physically meaningful quantity for any field line if we impose 
\begin{equation}
{\bf n}\times{\bf A}|_{\partial V} = {\bf n}\times{\bf A}^{\rm ref}|_{\partial V},
\label{eqn:gcon}
\end{equation}
where ${\bf A}^{\rm ref}=(r/2){\bf e}_\phi$ is the vector potential of a reference field ${\bf B}^{\rm ref}={\bf e}_z$ that matches ${\bf B}$ on $\partial V$. A similar restriction is used to ensure gauge invariance in the well-known relative magnetic helicity \cite{berger1984,*finn1985}. To demonstrate how ${\cal A}$ becomes meaningful, define a ``poloidal surface'' with flux $\Phi(\phi)$ bounded by (i) the field line, (ii) a vertical line on the side boundary $r=R$ at azimuth $\phi$, (iii) a straight line $L_0$ on $z=0$ joining the startpoints of the first two lines, and (iv) a straight line $L_1$ on $z=1$ joining their endpoints (Fig. \ref{fig:sketch}b). In view of the boundary condition ${\bf v}|_{\partial V}=0$, this is a comoving surface and $\Phi(\phi)$ is an ideal invariant for any $\phi$. By Stokes' Theorem,
\begin{equation}
\Phi(\phi) = {\cal A}(x_0) + \int_{L_0+L_1}{\bf A}\cdot\,\mathrm{d}{\bf l},
\label{eqn:phiphi}
\end{equation}
with no contribution from the side boundary since $A_z=0$ there by our gauge choice. For a periodic field line, the integrals in \eqref{eqn:phiphi} along $L_0$ and $L_1$ will be equal and opposite, so the flux $\Phi(\phi)$ is independent of the ``viewing angle'' $\phi$ and is given by ${\cal A}(x_0)$. For a non-periodic field line, the integrals need not cancel, and will depend in general on $\phi$. However, we can show that in this case ${\cal A}(x_0)$ gives the average flux over all viewing angles. To see this, consider the shaded quadrilateral lying on the lower boundary in Fig. \ref{fig:sketch}(b). The integral over $L_0$ and $L_1$ returns the magnetic flux through this quadrilateral, since $A_r=0$ on $z=0,1$. Since $B_z=1$, this is simply the area of the quadrilateral,
\begin{equation}
\int_{L_0+L_1}{\bf A}\cdot\,\mathrm{d}{\bf l} = \frac{R}{2}\big\{F_r\sin(F_\phi - \phi) + r_0\sin(\phi_0 - \phi) \big\}.
\end{equation}
This expression vanishes on averaging over $\phi$ so, for this gauge restriction,
\begin{equation}
\frac{1}{2\pi}\int_0^{2\pi}\Phi(\phi)\,\mathrm{d}\phi = {\cal{A}}(x_0)
\end{equation}
for any field line.

A fundamental property of ${\cal A}$ in our restricted gauge \eqref{eqn:gcon} is ideal invariance. One can see this by calculating $\mathrm{d}{\cal A}/\mathrm{d}t$ as a line integral over a moving domain \cite{frankel1997}. In an ideal evolution, ${\partial{\bf A}}/{\partial t} = {\bf v}\times{\bf B} + \nabla\psi$, so
\begin{align}
\frac{\mathrm{d}{\cal A}}{\mathrm{d}t} &= \int_{z=0}^{z=1}\left\{\frac{\partial{\bf A}}{\partial t} - {\bf v}\times\nabla\times{\bf A} + \nabla({\bf v}\cdot{\bf A}) \right\}\cdot\,\mathrm{d}{\bf l}\\
&= \int_{z=0}^{z=1}\nabla(\psi + {\bf v}\cdot{\bf A})\cdot\,\mathrm{d}{\bf l}.
\end{align}
The second term vanishes since ${\bf v}|_{\partial V}=0$. Under our gauge restriction \eqref{eqn:gcon}, $\psi$ is constant on $\partial V$ and the integral vanishes.

\section{Physical Interpretations} \label{sec:interp}

Before proving our main result, we describe three illuminating interpretations of the topological flux function.

\subsection{Field Line Helicity}

There is a simple relation between ${\cal A}$ and the magnetic helicity\cite{berger1988}. Since $V$ is magnetically open, we use the relative helicity $H_r$ (Ref. \onlinecite{berger1984,*finn1985}). Under our gauge conditions, however, this reduces to the same expression as the total helicity,
\begin{equation}
H_r \equiv \int_V({\bf A} + {\bf A}^{\rm ref})\cdot({\bf B} - {\bf B}^{\rm ref})\,\mathrm{d}^3x = \int_V {\bf A}\cdot{\bf B}\,\mathrm{d}^3x.
\end{equation}
Now suppose we change variables from $(r,\phi,z)$ to $(r_0,\phi_0,z)$, where $(r,\phi,z)=f(x_0; z)$ and $x_0\equiv(r_0,\phi_0,0)$ is the footpoint on $z=0$ of the field line through $(r,\phi,z)$. The Jacobian of this coordinate transformation is
\begin{equation}
\det(J) = \frac{r_0B_z(x_0)}{rB_z(r,\phi,z)}
\end{equation}
as may be verified by considering a thin flux tube around the field line and using $\nabla\cdot{\bf B}=0$. Thus we can re-write
\begin{align}
H_r&=\int_V{\bf A}(r,\phi,z)\cdot{\bf B}(r,\phi,z)\,r\,\mathrm{d}r\mathrm{d}\phi \mathrm{d}z\\
&= \int{\bf A}\big(f(x_0;z)\big)\cdot{\bf B}\big(f(x_0;z)\big)\frac{B_z(x_0)}{B_z\big(f(x_0;z)\big)}\,\mathrm{d}^2x_0\mathrm{d}z\\
&=\int_{z=0}{\cal A}(x_0)B_z(x_0)\,\mathrm{d}^2x_0.
\label{eqn:ahr}
\end{align}
So ${\cal A}$ is a density for $H_r$ in the cross-sectional plane, weighted by magnetic flux. (With our boundary conditions, we simply have $B_z(x_0)=1$.) Hence Berger \cite{berger1988} calls ${\cal A}$ a ``field line helicity''. Clearly it is possible for ${\cal A}\neq 0$ even when $H_r=0$, providing that the integral of ${\cal A}$ over all field lines vanishes.

\begin{figure}
\includegraphics[width=\columnwidth]{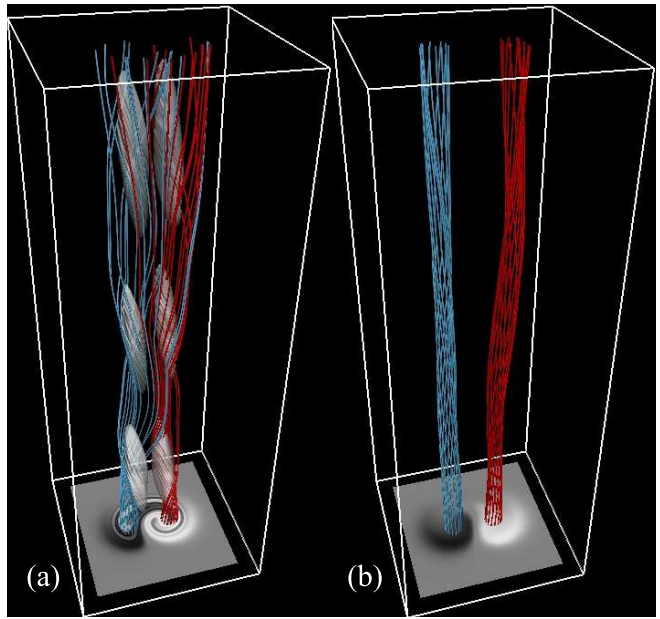}%
\caption{(Color online) Example calculation of the topological flux function ${\cal A}$ in a numerical magnetohydrodynamic simulation of magnetic relaxation \cite{wilmotsmith2010,*pontin2010}, using the gauge conditions \eqref{eqn:gcon} and $A_r=0$. Greyscale contours on the lower boundary show ${\cal A}$. Panels (a) and (b) correspond to before and after the relaxation respectively. \label{fig:sim}}%
\end{figure}

An example of this is the magnetohydrodynamic simulation of a relaxing solar coronal loop described by Wilmot-Smith {\it et al.}\cite{wilmotsmith2010,*pontin2010} and shown in Fig. \ref{fig:sim}. Although the evolution is resistive and there is widespread reconnection as the field relaxes, the relaxation is sufficiently fast to preserve the initial helicity $H_r=0$, as predicted by Taylor theory \cite{taylor1974}. But contrary to Taylor theory, which would predict a uniform relaxed field ${\bf B}={\bf e}_z$ with ${\cal A}\equiv 0$, the final state maintains equal regions of positive and negative ${\cal A}$, manifesting itself in non-trivial topology of the field lines in the end-state of the relaxation, despite the conservation of helicity \cite{yeates2010} (see also Ref. \onlinecite{candelaresi2011}). This illustrates how ${\cal A}$ contains more detailed information about the topology than $H_r$.

\subsection{Average Crossing Number}

The identification of ${\cal A}(x_0)$ with an average poloidal flux (Section \ref{sec:def}) suggests the following alternative topological interpretation. Given the field line $f(x_0;z)$ and a second field line $f(y_0;z)$, let $\theta_{x_0,y_0}(z)$ denote the orientation of the line segment connecting $f(x_0;z)$ and $f(y_0;z)$ in the plane at height $z$. Defining the signed crossing number\cite{berger1988} between these two field lines as the net winding angle
\begin{equation}
c_{x_0,y_0} = \frac{1}{2\pi}\int_0^1\frac{\mathrm{d}\theta_{x_0,y_0}(z)}{\mathrm{d}z}\,\mathrm{d}z,
\end{equation}
one can show (see the Appendix) that
\begin{equation}
{\cal A}(x_0) = \int_{z=0}c_{x_0,y_0}B_z(y_0)\,\mathrm{d}^2y_0.
\label{eqn:cross}
\end{equation}
In other words, ${\cal A}(x_0)$ is the average pairwise crossing number with all other field lines. In view of the relation \eqref{eqn:ahr}, this is consistent with Berger's formula\cite{berger1986}
\begin{equation}
H_r = \int_{z=0}\int_{z=0}c_{x_0,y_0}B_z(x_0)B_z(y_0)\,\mathrm{d}^2x_0\mathrm{d}^2y_0
\end{equation}
for the relative helicity of a magnetic field between two planes.

\subsection{Hamiltonian Action}

To motivate why ${\cal A}$ might be a sufficient condition for two magnetic braids to be topologically equivalent, we point out another physical interpretation of ${\cal A}$ that demonstrates a deep connection with the magnetic field structure. Namely, ${\cal A}$ is the action in a variational formulation that leads to the equations of the magnetic field lines \cite{cary1983}. In other words, for given vector potential ${\bf A}$ and field line mapping, the magnetic field lines ${\bf x}(l)$ are given by the Euler-Lagrange equations that extremize the integral
\begin{equation}
{\cal A}(x_0) = \int_{z=0}^{z=1}{\bf A}({\bf x})\cdot\frac{\mathrm{d}{\bf x}}{\mathrm{d}l}\,\mathrm{d}l.
\label{eqn:ham}
\end{equation}

It is well known that the magnetic field lines are trajectories of a Hamiltonian system---a fact well exploited in the modeling of toroidal fusion devices \cite{morrison2000,*caldas2012}. So the action \eqref{eqn:ham} is that of a Hamiltonian system, although it is not necessarily written in canonical coordinates \cite{cary1983}. To demonstrate that it really is Hamiltonian, we can re-write \eqref{eqn:ham} in canonical coordinates by a gauge transformation ${\bf A}\rightarrow{\bf A} + \nabla\chi$. This adds $\mathrm{d}\chi/\mathrm{d}l$ to the integrand (the Lagrangian), which leaves the field line equations (the Euler-Lagrange equations) unchanged. We can then write our system in canonical coordinates by choosing an appropriate gauge. If we set $A_r=0$ everywhere in space, then ${\bf A}\cdot\,\mathrm{d}{\bf l}=A_\phi\,r\,\mathrm{d}\phi + A_z\,\mathrm{d}z$. Making the identifications $p\leftrightarrow rA_\phi(r,\phi,z)$, $\quad q\leftrightarrow \phi$, $\quad t\leftrightarrow z$, $\quad H \leftrightarrow -A_z(r,\phi,z)$,
our action becomes
\begin{equation}
{\cal A} = \int_{0}^{1}\left(p\frac{\mathrm{d}q}{\mathrm{d}t} - H(p,q,t)\right)\,\mathrm{d}t.
\end{equation}
This is a 1 degree-of-freedom Hamiltonian system in canonical form. The generalized coordinate is $\phi$, the generalized momentum is $rA_\phi$, and the Hamiltonian is $-A_z$. Time corresponds to our $z$ coordinate, so our Hamiltonian is in general time dependent. We remark that the canonical gauge choice $A_r=0$ is equivalent to writing ${\bf B}$ in the form
${\bf B}=\nabla(rA_\phi)\times\nabla\phi + \nabla(A_z)\times\nabla z$,
which is widely used in toroidal plasmas \cite{boozer1983,*yoshida1994}. This gauge is also consistent with the gauge restriction \eqref{eqn:gcon} imposed to ensure ideal invariance of ${\cal A}$.

\section{Unique Topological Characterization} \label{sec:main}

Having identified ${\cal A}$ as the action in a Hamiltonian system, we now use general results about Hamiltonian systems to prove our main result that ${\cal A}$ is not only a necessary condition but also a sufficient condition for two magnetic braids to be topologically equivalent. The argument is based on work of Haro \cite{haro1998,*haro2000} on primitive functions of exact symplectomorphisms, and is best expressed in terms of differential forms. We assume that ${\bf A}$ is in canonical gauge $A_r=0$ and satisfies \eqref{eqn:gcon}.

As a special case of a more general result about Hamiltonian systems, it follows that
\begin{equation}
\mathrm{d}{\cal A} = F^*\alpha - \alpha,
\label{eqn:lemma}
\end{equation}
where $\alpha=(r^2/2)\,\mathrm{d}\phi$ is the canonical (Liouville) 1-form (Ref. \onlinecite{marsden1994}, p.148). For more details and proof of \eqref{eqn:lemma}, see, for example, Ref. \onlinecite{mcduff1995} (Proposition 9.18) or Haro \cite{haro1998,*haro2000}, where ${\cal A}$ is called a ``primitive function'' for $F$. Writing out the two components explicitly, \eqref{eqn:lemma} says that 
\begin{equation}
\frac{\partial{\cal A}}{\partial r_0} = \left(\frac{(F_r)^2}{2}\right)\frac{\partial F_\phi}{\partial r_0},\quad
\frac{\partial{\cal A}}{\partial \phi_0} = \left(\frac{(F_r)^2}{2}\right)\frac{\partial F_\phi}{\partial \phi_0} - \frac{r_0^2}{2}.
\label{eqn:pullback}
\end{equation}

To prove that ${\cal A}$ uniquely determines the topology, suppose that we have two magnetic braids ${\bf B}$, $\widetilde{\bf B}$, with topological flux functions ${\cal A}$, $\widetilde{\cal A}$ and field line mappings $F$, $\widetilde{F}$ respectively. We already know that $\widetilde{F}=F$ implies $\widetilde{\cal A}={\cal A}$, since ${\cal A}$ is an ideal invariant. But we can also see this from \eqref{eqn:lemma}, which gives
\begin{equation}
\mathrm{d}\widetilde{\cal A} = \widetilde{F}^*\alpha-\alpha=F^*\alpha-\alpha=\mathrm{d}{\cal A},
\end{equation}
so that $\widetilde{\cal A}$ and ${\cal A}$ differ by at most an overall constant, which vanishes since both braids satisfy ${\cal A}(R,\phi)=0$.

To prove the converse, assume that $\widetilde{\cal A}={\cal A}$, and define the mapping $G\equiv \widetilde{F}\circ F^{-1}$. Then, using \eqref{eqn:lemma},
\begin{align}
G^*\alpha-\alpha &= (F^{-1})^*\circ\widetilde{F}^*\alpha - \alpha\\
&= (F^{-1})^*(\alpha + \mathrm{d}{\cal A}) - \alpha\\
&= (F^{-1})^*\alpha - \alpha + (F^{-1})^*\mathrm{d}{\cal A}\\
&= (F^{-1})^*\alpha - (F^{-1})^*\circ F^* \alpha+ (F^{-1})^*\mathrm{d}{\cal A}\\
&= (F^{-1})^*(\alpha - F^*\alpha)+ (F^{-1})^*\mathrm{d}{\cal A}\\
&= -(F^{-1})^*\mathrm{d}{\cal A}+ (F^{-1})^*\mathrm{d}{\cal A}\\
&= 0.
\end{align}
Now we determine the possible mappings $G$ satisfying $G^*\alpha=\alpha$, or equivalently (compare Eq. \eqref{eqn:pullback})
\begin{equation}
\frac{(G_r)^2}{2}\frac{\partial G_\phi}{\partial r_0}\,\mathrm{d}r_0 + \left[\frac{(G_r)^2}{2}\frac{\partial G_\phi}{\partial \phi_0} - \frac{r_0^2}{2}\right]\,\mathrm{d}\phi_0 = 0.
\end{equation}
The $r_0$ and $\phi_0$ components give, respectively, $G_\phi=g(\phi_0)$ and $G_r=r_0(\mathrm{d}g/\mathrm{d}\phi_0)^{-1/2}$. But the possibilities are restricted by our boundary conditions: firstly, $G_r(R,\phi_0)=R$ implies that $\mathrm{d}g/\mathrm{d}\phi_0=1$, so that $G$ is a rigid rotation $G(r_0,\phi_0)=(r_0,\phi_0+\textrm{const})$.
But then $G_\phi(R,\phi_0)=\phi_0$ implies that $G={\rm id}$, and so $\widetilde{F}=F$. This completes the proof.

\section{Discussion} \label{sec:conclusion}

For clarity of presentation, we have assumed certain boundary conditions on the magnetic field, namely that ${\bf B}=1$ on all boundaries of our cylinder $V$, and that ${\bf v}=0$ on the bottom and top boundaries. We indicate here how the results generalize when these conditions are relaxed.

If one allows $B_\phi\neq 0$ on the side boundary $r=R$, then $G$ is determined only up to an overall rigid rotation, and we would have the result that $\widetilde{\cal A}$ and ${\cal A}$ differ by a constant if and only if $\widetilde{F}$ and $F$ differ by an overall rigid rotation. Such a rotation can be detected from knowledge of $F$ and $\widetilde{F}$ on the side boundary alone. In physical applications, the mapping on the side boundary may well be fixed in time.

It is also possible to consider more general $B_z$ distributions on the boundaries $z=0,1$, providing ${\bf A}$ satisfies \eqref{eqn:gcon} for an appropriate ${\bf A}^{\rm ref}$. In that case, the canonical 1-form $\alpha$ is $\alpha=rA_\phi(r,\phi,z)\,\mathrm{d}\phi$, and the possible mappings $G$ that preserve $\alpha$ (and are undetected by ${\cal A}$) may differ, although they take the general form of a cotangent lift $G(\phi_0,p)=\big(T(\phi_0),p(\mathrm{d}T/\mathrm{d}\phi_0)^{-1}\big)$, where $p=rA_\phi$ is the generalized momentum (see Ref. \onlinecite{marsden1994}, Proposition 6.3.2). If $\widetilde{F}=F$ on the side boundary $r=R$, it follows that $T(\phi_0)=\phi_0$ and hence again $G={\rm id}$.

Finally, one could consider a toroidal domain where ${\bf B}$ and ${\bf v}$ are periodic in $z$ and
remove the condition that ${\bf v}= 0$ on the boundaries $z=0,1$. In that case, the freedom to shuffle around field line endpoints on $z=0,1$ means that topological equivalence is a weaker notion, although it is certainly not true that any two field line mappings are equivalent, as would be the case if one had the freedom to apply independent motions on both boundaries. In the periodic case, the cross-section $z=0,1$ is no longer distinguished by the boundary conditions. Changing cross-section has the same effect on $F$ as an ideal evolution, so that the new mapping may be written $\widetilde{F}=S\circ F\circ S^{-1}$ for some field line mapping $S$. If ${\cal A}$ is the flux function for $F$, then one can show that the flux function $\widetilde{\cal A}$ for $\widetilde{F}$ is given by
\begin{equation}
\widetilde{\cal A} = (S^{-1})^*({\cal A} + F^*\chi - \chi),
\end{equation}
where $\chi$ is related to the mapping $S$ by $S^*\alpha - \alpha = \mathrm{d}\chi$ (see Ref. \onlinecite{haro1998}). The practical problem of determining whether two given topological flux functions are related in this way remains for further investigation.


%
%

%

\begin{acknowledgments}
This research was supported by the UK Science \& Technology Facilities Council grant ST/G002436/1 to the University of Dundee. ARY benefited from discussions during the `Braids \& Applications' workshop hosted by the Centro di Ricerca Matematica Ennio De Giorgi in June 2011. We thank the anonymous referee for suggesting the crossing number interpretation.
\end{acknowledgments}

\appendix*

\section{}

Equation \eqref{eqn:cross} may be established by expressing ${\bf A}(r,\phi,s)$ as a two-dimensional Biot-Savart integral. Write $x=(r,\phi,s)$, and let $y=(r',\phi',s)$ be another point at the same height. Then since $B_r=0$ on the side boundary we can write ${\bf B}=\nabla\times{\bf A}$ with
\begin{equation}
{\bf A}(x)=\frac{1}{2\pi}\int_{z=s}\frac{{\bf B}(y)\times{\bf r}}{|{\bf r}|^2}\,\mathrm{d}^2y,
\end{equation}
where ${\bf r}=x-y$ is the vector connecting $x$ and $y$, in the plane $z=s$. Notice that this satisfies the required gauge condition when $s=0$ or $s=1$.

Splitting ${\bf B}={\bf B}_\perp + B_z{\bf e}_z$, where ${\bf e}_z\cdot{\bf B}_\perp=0$, we find
\begin{equation}
\frac{{\bf A}(x)\cdot{\bf B}(x)}{B_z(x)} = \frac{1}{2\pi}\int_{z=s}\left(\frac{{\bf B}_\perp(x)}{B_z(x)} - \frac{{\bf B}_\perp(y)}{B_z(y)} \right)\cdot\frac{{\bf e}_z\times{\bf r}}{|{\bf r}|^2}B_z(y)\,\mathrm{d}^2y.
\end{equation}
Expressing $x=f(x_0;s)$ and $y=f(y_0;s)$ in terms of their associated footpoints $x_0$, $y_0$, $x=f(x_0;s)$ and $y=f(y_0;s)$ gives
\begin{equation}
\frac{{\bf A}\big(f(x_0;s)\big)\cdot{\bf B}\big(f(x_0;s)\big)}{B_z\big(f(x_0;s)\big)} = \frac{1}{2\pi}\int_{z=0}\frac{d\theta_{x_0,y_0}(s)}{dt}B_z(y_0)\,\mathrm{d}^2y_0.
\end{equation}
Integrating from $s=0$ to $s=1$ and using \eqref{eqn:adef} then establishes Equation \eqref{eqn:cross}. We note that Berger\cite{berger1988} derives a similar result for magnetic fields in a half space, but using a different argument where ${\cal A}(x_0)$ is considered as the limiting helicity of an infinitesimal flux tube around the field line $f(x_0;z)$.

\bibliography{yeates_braided}

\begin{thebibliography}{36}%
\makeatletter
\providecommand \@ifxundefined [1]{%
 \@ifx{#1\undefined}
}%
\providecommand \@ifnum [1]{%
 \ifnum #1\expandafter \@firstoftwo
 \else \expandafter \@secondoftwo
 \fi
}%
\providecommand \@ifx [1]{%
 \ifx #1\expandafter \@firstoftwo
 \else \expandafter \@secondoftwo
 \fi
}%
\providecommand \natexlab [1]{#1}%
\providecommand \enquote  [1]{``#1''}%
\providecommand \bibnamefont  [1]{#1}%
\providecommand \bibfnamefont [1]{#1}%
\providecommand \citenamefont [1]{#1}%
\providecommand \href@noop [0]{\@secondoftwo}%
\providecommand \href [0]{\begingroup \@sanitize@url \@href}%
\providecommand \@href[1]{\@@startlink{#1}\@@href}%
\providecommand \@@href[1]{\endgroup#1\@@endlink}%
\providecommand \@sanitize@url [0]{\catcode `\\12\catcode `\$12\catcode
  `\&12\catcode `\#12\catcode `\^12\catcode `\_12\catcode `\%12\relax}%
\providecommand \@@startlink[1]{}%
\providecommand \@@endlink[0]{}%
\providecommand \url  [0]{\begingroup\@sanitize@url \@url }%
\providecommand \@url [1]{\endgroup\@href {#1}{\urlprefix }}%
\providecommand \urlprefix  [0]{URL }%
\providecommand \Eprint [0]{\href }%
\providecommand \doibase [0]{http://dx.doi.org/}%
\providecommand \selectlanguage [0]{\@gobble}%
\providecommand \bibinfo  [0]{\@secondoftwo}%
\providecommand \bibfield  [0]{\@secondoftwo}%
\providecommand \translation [1]{[#1]}%
\providecommand \BibitemOpen [0]{}%
\providecommand \bibitemStop [0]{}%
\providecommand \bibitemNoStop [0]{.\EOS\space}%
\providecommand \EOS [0]{\spacefactor3000\relax}%
\providecommand \BibitemShut  [1]{\csname bibitem#1\endcsname}%
\let\auto@bib@innerbib\@empty
\bibitem [{\citenamefont {Longcope}(2005)}]{longcope2005}%
  \BibitemOpen
  \bibfield  {author} {\bibinfo {author} {\bibfnamefont {D.~W.}\ \bibnamefont
  {Longcope}},\ }\href@noop {} {\bibfield  {journal} {\bibinfo  {journal}
  {Living Rev. Solar Phys.}\ }\textbf {\bibinfo {volume} {2}},\ \bibinfo
  {pages} {7} (\bibinfo {year} {2005})}\BibitemShut {NoStop}%
\bibitem [{\citenamefont {{J.~P.~H. Goedbloed}}\ and\ \citenamefont {{S.
  Poedts}}(2004)}]{goedbloed2004}%
  \BibitemOpen
  \bibfield  {author} {\bibinfo {author} {\bibnamefont {{J.~P.~H. Goedbloed}}}\
  and\ \bibinfo {author} {\bibnamefont {{S. Poedts}}},\ }\href@noop {} {\emph
  {\bibinfo {title} {{Principles of Magnetohydrodynamics}}}}\ (\bibinfo
  {publisher} {Cambridge University Press},\ \bibinfo {address} {Cambridge},\
  \bibinfo {year} {2004})\BibitemShut {NoStop}%
\bibitem [{\citenamefont {Parker}(1972)}]{parker1972}%
  \BibitemOpen
  \bibfield  {author} {\bibinfo {author} {\bibfnamefont {E.~N.}\ \bibnamefont
  {Parker}},\ }\href@noop {} {\bibfield  {journal} {\bibinfo  {journal}
  {Astrophys. J,}\ }\textbf {\bibinfo {volume} {174}},\ \bibinfo {pages} {499}
  (\bibinfo {year} {1972})}\BibitemShut {NoStop}%
\bibitem [{\citenamefont {van Ballegooijen}(1985)}]{vanballegooijen1985}%
  \BibitemOpen
  \bibfield  {author} {\bibinfo {author} {\bibfnamefont {A.~A.}\ \bibnamefont
  {van Ballegooijen}},\ }\href@noop {} {\bibfield  {journal} {\bibinfo
  {journal} {Astrophys. J.}\ }\textbf {\bibinfo {volume} {298}},\ \bibinfo
  {pages} {421} (\bibinfo {year} {1985})}\BibitemShut {NoStop}%
\bibitem [{\citenamefont {{Z. Miki\'{c}}}, \citenamefont {{D. D. Schnack}},\
  and\ \citenamefont {{G. van Hoven}}(1989)}]{mikic1989}%
  \BibitemOpen
  \bibfield  {author} {\bibinfo {author} {\bibnamefont {{Z. Miki\'{c}}}},
  \bibinfo {author} {\bibnamefont {{D. D. Schnack}}}, \ and\ \bibinfo {author}
  {\bibnamefont {{G. van Hoven}}},\ }\href@noop {} {\bibfield  {journal}
  {\bibinfo  {journal} {Astrophys. J.}\ }\textbf {\bibinfo {volume} {338}},\
  \bibinfo {pages} {1148} (\bibinfo {year} {1989})}\BibitemShut {NoStop}%
\bibitem [{\citenamefont {{D. W. Longcope}}\ and\ \citenamefont {{H. R.
  Strauss}}(1994)}]{longcope1994}%
  \BibitemOpen
  \bibfield  {author} {\bibinfo {author} {\bibnamefont {{D. W. Longcope}}}\
  and\ \bibinfo {author} {\bibnamefont {{H. R. Strauss}}},\ }\href@noop {}
  {\bibfield  {journal} {\bibinfo  {journal} {Astrophys. J.}\ }\textbf
  {\bibinfo {volume} {437}},\ \bibinfo {pages} {851} (\bibinfo {year}
  {1994})}\BibitemShut {NoStop}%
\bibitem [{\citenamefont {{K. Galsgaard}}\ and\ \citenamefont {{A.
  Nordlund}}(1996)}]{galsgaard1996}%
  \BibitemOpen
  \bibfield  {author} {\bibinfo {author} {\bibnamefont {{K. Galsgaard}}}\ and\
  \bibinfo {author} {\bibnamefont {{A. Nordlund}}},\ }\href@noop {} {\bibfield
  {journal} {\bibinfo  {journal} {J Geophys. Res. A}\ }\textbf {\bibinfo
  {volume} {101}},\ \bibinfo {pages} {13445} (\bibinfo {year}
  {1996})}\BibitemShut {NoStop}%
\bibitem [{\citenamefont {{P. K. Browning}}\ \emph {et~al.}(2008)\citenamefont
  {{P. K. Browning}}, \citenamefont {{C. Gerrard}}, \citenamefont {{A. W.
  Hood}}, \citenamefont {{R. Kevis}},\ and\ \citenamefont {{R. A. M. van der
  Linden}}}]{browning2008}%
  \BibitemOpen
  \bibfield  {author} {\bibinfo {author} {\bibnamefont {{P. K. Browning}}},
  \bibinfo {author} {\bibnamefont {{C. Gerrard}}}, \bibinfo {author}
  {\bibnamefont {{A. W. Hood}}}, \bibinfo {author} {\bibnamefont {{R. Kevis}}},
  \ and\ \bibinfo {author} {\bibnamefont {{R. A. M. van der Linden}}},\
  }\href@noop {} {\bibfield  {journal} {\bibinfo  {journal} {Astron.
  Astrophys.}\ }\textbf {\bibinfo {volume} {485}},\ \bibinfo {pages} {837}
  (\bibinfo {year} {2008})}\BibitemShut {NoStop}%
\bibitem [{\citenamefont {{D. I. Pontin}}\ \emph {et~al.}(2010)\citenamefont
  {{D. I. Pontin}}, \citenamefont {{A. L. Wilmot-Smith}}, \citenamefont {{G.
  Hornig}},\ and\ \citenamefont {{K. Galsgaard}}}]{pontin2010}%
  \BibitemOpen
  \bibfield  {author} {\bibinfo {author} {\bibnamefont {{D. I. Pontin}}},
  \bibinfo {author} {\bibnamefont {{A. L. Wilmot-Smith}}}, \bibinfo {author}
  {\bibnamefont {{G. Hornig}}}, \ and\ \bibinfo {author} {\bibnamefont {{K.
  Galsgaard}}},\ }\href@noop {} {\bibfield  {journal} {\bibinfo  {journal}
  {Astron. Astrophys.}\ }\textbf {\bibinfo {volume} {525}},\ \bibinfo {pages}
  {A57} (\bibinfo {year} {2010})}\BibitemShut {NoStop}%
\bibitem [{\citenamefont {Morrison}(2000)}]{morrison2000}%
  \BibitemOpen
  \bibfield  {author} {\bibinfo {author} {\bibfnamefont {P.~J.}\ \bibnamefont
  {Morrison}},\ }\href@noop {} {\bibfield  {journal} {\bibinfo  {journal}
  {Phys. Plasmas}\ }\textbf {\bibinfo {volume} {7}},\ \bibinfo {pages} {2279}
  (\bibinfo {year} {2000})}\BibitemShut {NoStop}%
\bibitem [{\citenamefont {{I.~L. Caldas}}\ \emph {et~al.}(2012)\citenamefont
  {{I.~L. Caldas}}, \citenamefont {{R.~L. Viana}}, \citenamefont {{J.~D.
  Szezech}}, \citenamefont {{J.~S.~E. Portela}}, \citenamefont {{J. Fonseca}},
  \citenamefont {{M. Roberto}}, \citenamefont {{C.~G.~L. Martins}},\ and\
  \citenamefont {{E.~J. da Silva}}}]{caldas2012}%
  \BibitemOpen
  \bibfield  {author} {\bibinfo {author} {\bibnamefont {{I.~L. Caldas}}},
  \bibinfo {author} {\bibnamefont {{R.~L. Viana}}}, \bibinfo {author}
  {\bibnamefont {{J.~D. Szezech}}}, \bibinfo {author} {\bibnamefont {{J.~S.~E.
  Portela}}}, \bibinfo {author} {\bibnamefont {{J. Fonseca}}}, \bibinfo
  {author} {\bibnamefont {{M. Roberto}}}, \bibinfo {author} {\bibnamefont
  {{C.~G.~L. Martins}}}, \ and\ \bibinfo {author} {\bibnamefont {{E.~J. da
  Silva}}},\ }\href@noop {} {\bibfield  {journal} {\bibinfo  {journal} {Comm.
  Nonlin. Sci. Num. Simulat.}\ }\textbf {\bibinfo {volume} {17}},\ \bibinfo
  {pages} {2021} (\bibinfo {year} {2012})}\BibitemShut {NoStop}%
\bibitem [{\citenamefont {{V. S. Titov}}, \citenamefont {{G. Hornig}},\ and\
  \citenamefont {{P. D\'{e}moulin}}(2002)}]{titov2002}%
  \BibitemOpen
  \bibfield  {author} {\bibinfo {author} {\bibnamefont {{V. S. Titov}}},
  \bibinfo {author} {\bibnamefont {{G. Hornig}}}, \ and\ \bibinfo {author}
  {\bibnamefont {{P. D\'{e}moulin}}},\ }\href@noop {} {\bibfield  {journal}
  {\bibinfo  {journal} {J. Geophys. Res. A}\ }\textbf {\bibinfo {volume}
  {107}},\ \bibinfo {pages} {1164} (\bibinfo {year} {2002})}\BibitemShut
  {NoStop}%
\bibitem [{\citenamefont {Titov}(2007)}]{titov2007}%
  \BibitemOpen
  \bibfield  {author} {\bibinfo {author} {\bibfnamefont {V.~S.}\ \bibnamefont
  {Titov}},\ }\href@noop {} {\bibfield  {journal} {\bibinfo  {journal}
  {Astrophys. J.}\ }\textbf {\bibinfo {volume} {660}},\ \bibinfo {pages} {863}
  (\bibinfo {year} {2007})}\BibitemShut {NoStop}%
\bibitem [{\citenamefont {{A. R. Yeates}}, \citenamefont {{G. Hornig}},\ and\
  \citenamefont {{B. T. Welsch}}(2012)}]{yeates2012}%
  \BibitemOpen
  \bibfield  {author} {\bibinfo {author} {\bibnamefont {{A. R. Yeates}}},
  \bibinfo {author} {\bibnamefont {{G. Hornig}}}, \ and\ \bibinfo {author}
  {\bibnamefont {{B. T. Welsch}}},\ }\href@noop {} {\bibfield  {journal}
  {\bibinfo  {journal} {Astron. Astrophys.}\ }\textbf {\bibinfo {volume}
  {539}},\ \bibinfo {pages} {A1} (\bibinfo {year} {2012})}\BibitemShut
  {NoStop}%
\bibitem [{\citenamefont {{E. Pariat}}\ and\ \citenamefont {{P.
  D\'{e}moulin}}(2012)}]{pariat2012}%
  \BibitemOpen
  \bibfield  {author} {\bibinfo {author} {\bibnamefont {{E. Pariat}}}\ and\
  \bibinfo {author} {\bibnamefont {{P. D\'{e}moulin}}},\ }\href@noop {}
  {\bibfield  {journal} {\bibinfo  {journal} {Astrophys. J.}\ }\textbf
  {\bibinfo {volume} {541}},\ \bibinfo {pages} {A78} (\bibinfo {year}
  {2012})}\BibitemShut {NoStop}%
\bibitem [{\citenamefont {{Brown}}, \citenamefont {{Canfield}},\ and\
  \citenamefont {{Pevtsov}}(1999)}]{brown1999}%
  \BibitemOpen
  \bibinfo {editor} {\bibfnamefont {M.~R.}\ \bibnamefont {{Brown}}}, \bibinfo
  {editor} {\bibfnamefont {R.~C.}\ \bibnamefont {{Canfield}}}, \ and\ \bibinfo
  {editor} {\bibfnamefont {A.~A.}\ \bibnamefont {{Pevtsov}}},\ eds.,\
  \href@noop {} {\emph {\bibinfo {title} {{Magnetic Helicity in Space and
  Laboratory Plasmas, Geophysical Monograph 111}}}}\ (\bibinfo  {publisher}
  {AGU},\ \bibinfo {year} {1999})\BibitemShut {NoStop}%
\bibitem [{\citenamefont {Taylor}(1974)}]{taylor1974}%
  \BibitemOpen
  \bibfield  {author} {\bibinfo {author} {\bibfnamefont {J.~B.}\ \bibnamefont
  {Taylor}},\ }\href@noop {} {\bibfield  {journal} {\bibinfo  {journal} {Phys.
  Rev. Lett.}\ }\textbf {\bibinfo {volume} {33}},\ \bibinfo {pages} {1139}
  (\bibinfo {year} {1974})}\BibitemShut {NoStop}%
\bibitem [{\citenamefont {Taylor}(2000)}]{taylor2000}%
  \BibitemOpen
  \bibfield  {author} {\bibinfo {author} {\bibfnamefont {J.~B.}\ \bibnamefont
  {Taylor}},\ }\href@noop {} {\bibfield  {journal} {\bibinfo  {journal} {Phys.
  Plasmas}\ }\textbf {\bibinfo {volume} {7}},\ \bibinfo {pages} {1623}
  (\bibinfo {year} {2000})}\BibitemShut {NoStop}%
\bibitem [{\citenamefont {Berger}(1988)}]{berger1988}%
  \BibitemOpen
  \bibfield  {author} {\bibinfo {author} {\bibfnamefont {M.~A.}\ \bibnamefont
  {Berger}},\ }\href@noop {} {\bibfield  {journal} {\bibinfo  {journal}
  {Astron. Astrophys.}\ }\textbf {\bibinfo {volume} {201}},\ \bibinfo {pages}
  {355} (\bibinfo {year} {1988})}\BibitemShut {NoStop}%
\bibitem [{\citenamefont {{J. R. Cary}}\ and\ \citenamefont {{R. G.
  Littlejohn}}(1983)}]{cary1983}%
  \BibitemOpen
  \bibfield  {author} {\bibinfo {author} {\bibnamefont {{J. R. Cary}}}\ and\
  \bibinfo {author} {\bibnamefont {{R. G. Littlejohn}}},\ }\href@noop {}
  {\bibfield  {journal} {\bibinfo  {journal} {Annal. Phys.}\ }\textbf {\bibinfo
  {volume} {151}},\ \bibinfo {pages} {1} (\bibinfo {year} {1983})}\BibitemShut
  {NoStop}%
\bibitem [{\citenamefont {Petrisor}(2002)}]{petrisor2002}%
  \BibitemOpen
  \bibfield  {author} {\bibinfo {author} {\bibfnamefont {E.}~\bibnamefont
  {Petrisor}},\ }\href@noop {} {\bibfield  {journal} {\bibinfo  {journal}
  {Chaos Solitons Fractals}\ }\textbf {\bibinfo {volume} {14}},\ \bibinfo
  {pages} {117} (\bibinfo {year} {2002})}\BibitemShut {NoStop}%
\bibitem [{\citenamefont {{A. Apte}}, \citenamefont {{R. de la Llave}},\ and\
  \citenamefont {{E. Petrisor}}(2006)}]{apte2006}%
  \BibitemOpen
  \bibfield  {author} {\bibinfo {author} {\bibnamefont {{A. Apte}}}, \bibinfo
  {author} {\bibnamefont {{R. de la Llave}}}, \ and\ \bibinfo {author}
  {\bibnamefont {{E. Petrisor}}},\ }\href@noop {} {\bibfield  {journal}
  {\bibinfo  {journal} {Chaos Solitons Fractals}\ }\textbf {\bibinfo {volume}
  {27}},\ \bibinfo {pages} {1115} (\bibinfo {year} {2006})}\BibitemShut
  {NoStop}%
\bibitem [{\citenamefont {{A. R. Yeates}}\ and\ \citenamefont {{G.
  Hornig}}(2011)}]{yeates2011}%
  \BibitemOpen
  \bibfield  {author} {\bibinfo {author} {\bibnamefont {{A. R. Yeates}}}\ and\
  \bibinfo {author} {\bibnamefont {{G. Hornig}}},\ }\href@noop {} {\bibfield
  {journal} {\bibinfo  {journal} {Phys. Plasmas}\ }\textbf {\bibinfo {volume}
  {18}},\ \bibinfo {pages} {102118} (\bibinfo {year} {2011})}\BibitemShut
  {NoStop}%
\bibitem [{\citenamefont {{M. A. Berger}}\ and\ \citenamefont {{G. B.
  Field}}(1984)}]{berger1984}%
  \BibitemOpen
  \bibfield  {author} {\bibinfo {author} {\bibnamefont {{M. A. Berger}}}\ and\
  \bibinfo {author} {\bibnamefont {{G. B. Field}}},\ }\href@noop {} {\bibfield
  {journal} {\bibinfo  {journal} {J. Fluid Mech.}\ }\textbf {\bibinfo {volume}
  {147}},\ \bibinfo {pages} {133} (\bibinfo {year} {1984})}\BibitemShut
  {NoStop}%
\bibitem [{\citenamefont {{J. H. Finn}}\ and\ \citenamefont {{T. M.
  Antonsen}}(1985)}]{finn1985}%
  \BibitemOpen
  \bibfield  {author} {\bibinfo {author} {\bibnamefont {{J. H. Finn}}}\ and\
  \bibinfo {author} {\bibnamefont {{T. M. Antonsen}}},\ }\href@noop {}
  {\bibfield  {journal} {\bibinfo  {journal} {Comments Plasma Phys. Contr.
  Fusion}\ }\textbf {\bibinfo {volume} {9}},\ \bibinfo {pages} {111} (\bibinfo
  {year} {1985})}\BibitemShut {NoStop}%
\bibitem [{\citenamefont {Frankel}(1997)}]{frankel1997}%
  \BibitemOpen
  \bibfield  {author} {\bibinfo {author} {\bibfnamefont {T.}~\bibnamefont
  {Frankel}},\ }\href@noop {} {\emph {\bibinfo {title} {{The Geometry of
  Physics}}}}\ (\bibinfo  {publisher} {Cambridge University Press},\ \bibinfo
  {address} {Cambridge},\ \bibinfo {year} {1997})\ p.\ \bibinfo {pages}
  {143}\BibitemShut {NoStop}%
\bibitem [{\citenamefont {{A.~L. Wilmot-Smith}}, \citenamefont {{D.~I.
  Pontin}},\ and\ \citenamefont {{G. Hornig}}(2010)}]{wilmotsmith2010}%
  \BibitemOpen
  \bibfield  {author} {\bibinfo {author} {\bibnamefont {{A.~L. Wilmot-Smith}}},
  \bibinfo {author} {\bibnamefont {{D.~I. Pontin}}}, \ and\ \bibinfo {author}
  {\bibnamefont {{G. Hornig}}},\ }\href@noop {} {\bibfield  {journal} {\bibinfo
   {journal} {Astron. Astrophys.}\ }\textbf {\bibinfo {volume} {516}},\
  \bibinfo {pages} {A5} (\bibinfo {year} {2010})}\BibitemShut {NoStop}%
\bibitem [{\citenamefont {{A. R. Yeates}}, \citenamefont {{G. Hornig}},\ and\
  \citenamefont {{A. L. Wilmot-Smith}}(2010)}]{yeates2010}%
  \BibitemOpen
  \bibfield  {author} {\bibinfo {author} {\bibnamefont {{A. R. Yeates}}},
  \bibinfo {author} {\bibnamefont {{G. Hornig}}}, \ and\ \bibinfo {author}
  {\bibnamefont {{A. L. Wilmot-Smith}}},\ }\href@noop {} {\bibfield  {journal}
  {\bibinfo  {journal} {Phys. Rev. Lett.}\ }\textbf {\bibinfo {volume} {105}},\
  \bibinfo {pages} {085002} (\bibinfo {year} {2010})}\BibitemShut {NoStop}%
\bibitem [{\citenamefont {{S. Candelaresi}}\ and\ \citenamefont {{A.
  Brandenburg}}(2011)}]{candelaresi2011}%
  \BibitemOpen
  \bibfield  {author} {\bibinfo {author} {\bibnamefont {{S. Candelaresi}}}\
  and\ \bibinfo {author} {\bibnamefont {{A. Brandenburg}}},\ }\href@noop {}
  {\bibfield  {journal} {\bibinfo  {journal} {Phys. Rev. E}\ }\textbf {\bibinfo
  {volume} {84}},\ \bibinfo {pages} {01646} (\bibinfo {year}
  {2011})}\BibitemShut {NoStop}%
\bibitem [{\citenamefont {Berger}(1986)}]{berger1986}%
  \BibitemOpen
  \bibfield  {author} {\bibinfo {author} {\bibfnamefont {M.~A.}\ \bibnamefont
  {Berger}},\ }\href@noop {} {\bibfield  {journal} {\bibinfo  {journal}
  {Geophys. Astrophys. Fluid Dyn.}\ }\textbf {\bibinfo {volume} {34}},\
  \bibinfo {pages} {265} (\bibinfo {year} {1986})}\BibitemShut {NoStop}%
\bibitem [{\citenamefont {Boozer}(1983)}]{boozer1983}%
  \BibitemOpen
  \bibfield  {author} {\bibinfo {author} {\bibfnamefont {A.~H.}\ \bibnamefont
  {Boozer}},\ }\href@noop {} {\bibfield  {journal} {\bibinfo  {journal} {Phys.
  Fluids}\ }\textbf {\bibinfo {volume} {26}},\ \bibinfo {pages} {1288}
  (\bibinfo {year} {1983})}\BibitemShut {NoStop}%
\bibitem [{\citenamefont {Yoshida}(1994)}]{yoshida1994}%
  \BibitemOpen
  \bibfield  {author} {\bibinfo {author} {\bibfnamefont {Z.}~\bibnamefont
  {Yoshida}},\ }\href@noop {} {\bibfield  {journal} {\bibinfo  {journal} {Phys.
  Plasmas}\ }\textbf {\bibinfo {volume} {1}},\ \bibinfo {pages} {208} (\bibinfo
  {year} {1994})}\BibitemShut {NoStop}%
\bibitem [{\citenamefont {Haro}(1998)}]{haro1998}%
  \BibitemOpen
  \bibfield  {author} {\bibinfo {author} {\bibfnamefont {A.}~\bibnamefont
  {Haro}},\ }\href@noop {} {Ph.D. thesis},\ \bibinfo  {school} {Universitat de
  Barcelona}, \bibinfo {address} {Barcelona} (\bibinfo {year}
  {1998})\BibitemShut {NoStop}%
\bibitem [{\citenamefont {Haro}(2000)}]{haro2000}%
  \BibitemOpen
  \bibfield  {author} {\bibinfo {author} {\bibfnamefont {A.}~\bibnamefont
  {Haro}},\ }\href@noop {} {\bibfield  {journal} {\bibinfo  {journal}
  {Nonlinearity}\ }\textbf {\bibinfo {volume} {13}},\ \bibinfo {pages} {1483}
  (\bibinfo {year} {2000})}\BibitemShut {NoStop}%
\bibitem [{\citenamefont {{J. E. Marsden}}\ and\ \citenamefont {{T. S.
  Ratiu}}(1994)}]{marsden1994}%
  \BibitemOpen
  \bibfield  {author} {\bibinfo {author} {\bibnamefont {{J. E. Marsden}}}\ and\
  \bibinfo {author} {\bibnamefont {{T. S. Ratiu}}},\ }\href@noop {} {\emph
  {\bibinfo {title} {{Introduction to Mechanics and Symmetry}}}}\ (\bibinfo
  {publisher} {Springer-Verlag},\ \bibinfo {address} {New York},\ \bibinfo
  {year} {1994})\BibitemShut {NoStop}%
\bibitem [{\citenamefont {{D. McDuff}}\ and\ \citenamefont {{D.
  Salamon}}(1995)}]{mcduff1995}%
  \BibitemOpen
  \bibfield  {author} {\bibinfo {author} {\bibnamefont {{D. McDuff}}}\ and\
  \bibinfo {author} {\bibnamefont {{D. Salamon}}},\ }\href@noop {} {\emph
  {\bibinfo {title} {{Introduction to Symplectic Topology}}}}\ (\bibinfo
  {publisher} {Oxford University Press},\ \bibinfo {address} {Oxford},\
  \bibinfo {year} {1995})\BibitemShut {NoStop}%
\end{thebibliography}%
\end{document}